\RequirePackage{fix-cm}

\documentclass[smallextended]{svjour3}

\smartqed  

\usepackage{graphicx}
\usepackage{amsmath,amssymb}

\usepackage{bm}
\usepackage{cite}
\usepackage{float}
\usepackage{multirow}
\usepackage{caption}
\usepackage{array}

\begin{document}

\title{Linear complexity over ${\mathbb{F}_{{q}}}$ and 2-adic complexity of a class of binary generalized cyclotomic sequences with low-value autocorrelation}



\author{Yan Wang         \and
        Xilin Han     \and
        Weiqiong Wang   \and
        Ziling Heng 
}


\institute{ Yan Wang \at
              School of Science, Xi'an University of Architecture and
Technology, Xi'an, 710055, China\\
              Tel.: 13519162272\\
              Fax: \\
              \email{wangyan@xauat.edu.cn}           
           \and
           Xilin Han \at
              School of Science, Xi'an University of Architecture and
Technology, Xi'an, 710055, China\\
              \email{hanxilin121@xauat.edu.cn}
           \and
           Weiqiong Wang  \at
              School of Science, Chang'an University , Xi'an, 710064, China\\
              \email{wqwang@chd.edu.cn}
           \and
           Ziling Heng \at
             School of Science, Chang'an University , Xi'an, 710064, China\\
             \email{zilingheng@163.com}
}

\date{Received: date / Accepted: date}

\maketitle

\begin{abstract}
A class of binary sequences with period $2p$ is constructed using generalized cyclotomic classes, and their linear complexity, minimal polynomial over ${\mathbb{F}_{{q}}}$ as well as 2-adic complexity are determined using Gauss period and group ring theory. The results show that the linear complexity of these sequences attains the maximum when $p\equiv \pm 1(\bmod~8)$ and is equal to {$p$+1} when $p\equiv \pm 3(\bmod~8)$ over extension field. Moreover, the 2-adic complexity of these sequences is maximum. According to Berlekamp-Massey(B-M) algorithm and the rational approximation algorithm(RAA), these sequences have quite good cryptographyic properties in the aspect of linear complexity and 2-adic complexity.
\end{abstract}

\section{Introduction}
\label{intro}
Pseudorandom sequences have wide applications in simulation, software testing, global positioning systems, ranging systems, codedivision multiple-access systems, radar systems, spread-spectrum communication systems, and stream ciphers. Many applications in communication require sequences have good autocorrelation and good crosscorrelation, while
many applications in cryptography require sequences have high linear complexity and 2-adic complexity. A pseudorandom sequence can be generated by a linear feedback shift register(LFSR), and also can be generated by a feedback with carry shift register(FCSR). After the B-M algorithm \cite{1} for LFSRs and RAA \cite{2} for FCSRs were presented, linear complexity and 2-adic complexity have been regarded as critical security criterias and both are required to be no less than one half of the period.

Sequences with high linear complexity can be constructed based on cyclotomic classes. Let $n \ge 2$ be a positive integer. A partition
$\{ {D_0},{D_1}, \cdots ,{D_{d - 1}}\} $ of $\mathbb{Z}_n^*$ is a family of sets with ${D_i} \cap {D_j} = \emptyset $ for all $i \ne j$, $ \cup _{i = 0}^{d - 1}{D_i} = \mathbb{Z}_n^*.$ If ${D_0}$ is a multiplicative subgroup of $\mathbb{Z}_n^*$, and there are elements ${g_1}, \cdots ,{g_{d - 1}}$ of $\mathbb{Z}_n^*$ such that ${D_i} = {g_i}{D_0}$ for all $i$, then ${D_i}$ is called generalized cyclotomic classes of order $d$ when $n$ is composite, and classical cyclotomic classes of order $d$ when $n$ is prime.
Generally, sequences based on classical cyclotomic classes are called classical cyclotomic sequences, and sequences based on generalized cyclotomic classes are called generalized cyclotomic sequences. There are lots of work on linear complexity of binary cyclotomic sequences based on Whiteman generalized cyclotomy and Ding-Helleseth generalized cyclotomy \cite{3}\cite{4}\cite{5}. Zhang et al.\cite{6} constructed two classes of binary generalized cyclotomic sequences with period $2{{p}^{m}}$ and showed that such sequences have high linear complexity over $\mathbb{F}_2$. Later, Ke et al.\cite{23} determined the linear complexity of sequences and showed that the linear complexity of these sequences attained the maximum over $\mathbb{Z}_2$. Xiao et al.\cite{7} presented a new class of binary cyclotomic sequences with period ${{p}^{2}}$ and determined the linear complexity of these sequences in the case of $f={{2}^{r}}$ for a positive integer $r$. These sequences are shown have large linear complexity when $p$ is a non-Wieferich prime over $\mathbb{F}_2$, where ${d_j} = \frac{{\varphi ({p^j})}}{e} = {p^{j - 1}}f,p = ef + 1$ and ${d_j}$ are the order of generalized cyclotomic classes with respect to ${p^j}$. Afterwards, Edemskiy et al.\cite{8} determined the linear complexity of sequences and extended the result to more general cases. Ouyang and Xie \cite{9} constructed two classes of such binary generalized cyclotomic sequences having high linear complexity with period $2{{p}^{m}}(m\ge 2)$ over $\mathbb{F}_{2^n}$ based on the work of Edemskiy et al.. And wang et al.\cite{10} studied a new class of binary generalized cyclotomic sequences having high linear complexity with period $2{{p}^ {m}}$ for arbitrary prime order $q$.

High 2-adic complexity sequences are also concerned by many researchers. Klapper \cite{2} proposed the concept of 2-adic complexity and pointed out that a prime period $m$-sequence has maximal 2-adic complexity. Moreover, Tian and Qi \cite{11} proved that the 2-adic complexity of all binary $m$-sequence is maximal. Then, Xiong et al.\cite{12} presented a new method to compute the 2-adic complexity of binary sequences by circulant matrix. Using this method, they proved that the 2-adic complexity of a twin-prime sequence with 2-level autocorrelation is maximal. Hu \cite{13}  proposed a simple method to compute the 2-adic complexity of any periodic binary sequence with ideal two-level autocorrelation, and his approach is associated with the autocorrelation function of the sequence. Sun et al.\cite{14} showed that the 2-adic complexity of modified Jacobi sequences of period $pq$ satisfy a lower bound $pq-p-q-1$ and then Holfer and Winterhof \cite{15} proved they can be maximal in the case of $\frac{q+1}{4}\le p\le 4q-1$. Recently, Yang and Feng \cite{16} determined the exact value of 2-adic complexity of the generalized binary sequences with period $pq$ of order 2, particularly, they improved the lower bounds presented in \cite{14} and \cite{15}, and gave the condition for the maxiumn value of 2-adic complexity. Qiang et al.\cite{17} studied the 2-adic complexity of two classes of binary sequences with interleaved structure and showed that the 2-adic complexity of such sequences is large enough to resist the attack of the rational approximation algorithm. Moreover, Jing et al.\cite{18} determined the autocorrelation distribution and 2-adic complexity of generalized cyclotomic binary sequences of order $2$ with period $pq$ in a way by using group ring theory and a version of quadratic Gaussian sum valued in group ring $R=\mathbb{Z}[$G$]$ where $G$ is a cyclic group of order $n$.

In this paper, we determine the linear complexity and minimal polynomial of a class of generalized cyclotomic binary sequences with low-valued autocorrelation by explicitly computing the number of zeros of the generating polynomial $S(x)$ over ${\mathbb{F}_{q}}(q={{r}^{m}},r\ne p,r\ge 5)$. Further more, we show that the 2-adic complexity of these sequences by means of group ring operations. The rest of the paper is organized as follows. Section $2$ introduces some necessary preliminaries of our research. Section $3$ calculates the linear complexity and minimal polynomial of these sequences over a finite extension field ${\mathbb{F}_{q}}$, while the 2-adic complexity of these sequences is determined in section $4$. Section $5$ concludes this paper.

\section{Preliminaries}
\label{sec:1}
\subsection{Generalized cyclotomic sequences}
\label{sec:2}
Let $p$ be an odd prime and $g$ be an odd common primitive element of both $p$ and $2p$, $<{{g}^{2}}>$ denote the subgroup generated by $g^2$. Let ${\mathbb{Z}_{2p}}=\{0,1,2,\cdots ,2p-1\}$ denote the residue class ring of module $2p$ and $\mathbb {Z}_{2p}^*$ be the multiplicative group consisting of all invertible elements in $\mathbb {Z}_{2p}$.

Denote
    \begin{align*}
	    D_{0}^{(p)} & = <{{g}^{2}}>(\bmod p) \\
	& = \left\{ {{g}^{2k}}(\bmod~p):k=0,1,\cdots ,\frac{p-1}{2}-1 \right\},\\
	\end{align*}
    \begin{align*}
	    D_{0}^{(2p)} & = <{{g}^{2}}>(\bmod~2p) \\
	& = \left\{ {{g}^{2k}}(\bmod~2p):k=0,1,\cdots ,\frac{p-1}{2}-1 \right\},\\
	    D_{1}^{(p)} & = gD_{0}^{(p)}=g<{{g}^{2}}>(\bmod~p) \\
	& = \left\{ {{g}^{2k+1}}(\bmod~p):k=0,1,\cdots ,\frac{p-1}{2}-1 \right\},\\
	    D_{1}^{(2p)} & = gD_{0}^{(2p)}=g<{{g}^{2}}>(\bmod~2p) \\
	& = \left\{ {{g}^{2k+1}}(\bmod~2p):k=0,1,\cdots ,\frac{p-1}{2}-1 \right\},
	\end{align*}
where $D_0^{(n)}$ and $D_1^{(n)}$ are called general cyclotomic class of order $2$ with respect to $n$, then
    \begin{align*}
   \mathbb{Z}_{2p}^* & = D_0^{(2p)} \cup D_1^{(2p)}, \\
      \mathbb{Z}_p^* & = D_0^{(p)} \cup D_1^{(p)}, \\
      \mathbb{Z}_2^* & = \{ 1\},
	\end{align*}
moreover,
$${\mathbb{Z}_{2p}} = D_0^{(2p)} \cup D_1^{(2p)} \cup 2D_0^{(p)} \cup 2D_1^{(p)} \cup p\mathbb{Z}_2^* \cup \{ 0\} ,$$
where $\mathbb{Z}_n^* = \{ 1 \le i \le n - 1:\gcd (i,n) = 1\}.$

Denote
\[{C_0} = D_0^{(2p)} \cup 2D_0^{(p)} \cup p\mathbb{Z}_2^*,\]
\[{C_1} = D_1^{(2p)} \cup 2D_1^{(p)} \cup \left\{ 0 \right\},\]
then,
${\mathbb{Z}_{2p}} = {C_0} \cup {C_1}$ and ${C_0} \cap {C_1} = \emptyset $, where $\emptyset $ denotes the empty set.

 We consider a class of generalized cyclotomic  binary sequences with period $2p$ defined as
\begin{equation}
{s_i} = \left\{ {\begin{array}{*{20}{c}}
{0,}&{i(\bmod~{2p}) \in {C_0},}\\
{1,}&{i(\bmod~{2p}) \in {C_1}.}
\end{array}} \right.
\end{equation}

 Using cyclotomic numbers, we can determine such sequences have low value autocorrelation when $p \equiv 1(\bmod~8)$ and $p \equiv 3(\bmod~8)$. The autocorrelation spectrum of $\{ s_i\}$ is given in Table  \ref{table1}.

\begin{table}[h]
  \centering
  \caption{ The autocorrelation spectrum }\label{table1}
\begin{tabular}{|c|c|>{\centering\arraybackslash}m{3cm}|}
\hline
 $p$ & Value of $C_S(w)$ & Number of times given value occurs\\
  \hline
 \multirow{3}*{$p \equiv 1(\bmod~8)$} & $2p$ & $1$ \\ \cline{2-3}
                     & $2p-4$ & $1$  \\ \cline{2-3}
                     & $-2$    & $2p-2$  \\ \hline
 \multirow{4}*{$p \equiv 3(\bmod~8)$} & $2p$ & $1$ \\ \cline{2-3}
                     & $-2p$ & $1$  \\ \cline{2-3}
                     & $2$    & $p-1$  \\ \cline{2-3}
                     & $-2$    & $p-1$  \\ \hline

\end{tabular}
\end{table}

In this correspondence, we focus on the properties in the aspect of linear complexity and 2-adic complexity according to B-M algorithm and RAA.

\subsection{The linear complexity of periodic binary sequences over extention field}
\label{sec:2}
Let $p$ be an odd prime and $m$ is a positive integer, $r$ be an odd prime, $r \ne p,r \ge 5$, then ${\mathbb{F}_{{r^m}}}$ is an extension field of ${r^m}$ elements with characteristic $r$. Let $\{ s(t)\}(t = 0,1, \cdots ,N - 1) $ be a sequence with period $N$. The linear complexity of $\{ s(t)\} $ is the length of the shortest feedback shift register generated, that is, the smallest positive integer $L$ satisfies the following recurrent relation
$$s(t + L) = {c_{L - 1}}s(t + L - 1) +  \cdots  + {c_1}s(t + 1) + {c_0}s(t)$$
for $t \ge 0,{c_0},{c_1}, \cdots ,{c_{L - 1}} \in {\mathbb{F}_{{r^m}}}.$ The linear complexity of $\{ s(t)\} $ is denoted by $LC(s)$. The minimal polynomial related to $\{ s(t)\} $ is
\[m(x) = {x^L} - \sum\limits_{i = 0}^{L - 1} {{c_i}} {x^i}.\]
The generating polynomial related to $\{ s(t)\} $ is
\begin{equation}
S(x) = \sum\limits_{t = 0}^{N - 1} {s(t)} {x^t}.
\end{equation}
The following equation relates both the minimal polynomial and the generating polynomial of the sequences $\{ s(t)\} $
\begin{equation}
m(x) = \frac{{{x^N} - 1}}{{\gcd ({x^N} - 1,S(x))}}.
\end{equation}
Moerover, the linear complexity of $\{ s(t)\} $ can also be given by
\begin{equation}
LC(s) = \deg (m(x)) = N - \deg (\gcd ({x^N} - 1,S(x))).
\end{equation}

Let $m = or{d_p}(r),r \ne p,\beta $ be the $2{p^{th}}$ root of unity, according to formula $(4)$, we have
\begin{equation}
LC(s) = N - |\{ k:S({\beta ^k}) = 0,0 \le k < N\} |.
\end{equation}
\subsection{2-Adic complexity of periodic binary sequences}
\label{sec:2}
Any infinite binary sequence $s = ({s_0},{s_1},{s_2}, \cdots )$ may be identified with the formal power series $\alpha  = \sum\limits_{i = 0}^\infty  {{s_i}{2^i}} $. The collection of all such power series forms a ring, denoted by $\textbf{Z}_2$. There is a one to one correspondence between rational numbers $\alpha  = c/b$ (where $b$ is odd) and ultimately periodic binary sequences. Thus, the ring $\textbf{Z}_2$ contains all the rational numbers with odd denominator \cite{2}.

It turns out that the output sequences of FCSR are ultimately periodic sequences. Conversely, an ultimately periodic binary sequences corresponding to a rational number $\alpha  = c/b$, where $b$ is an odd positive integer, can be generated by an FCSR with connection integer $b$. In particular, $s$ is a strictly periodic sequence if and only if its corresponding rational number $\alpha $ satisfies that $ - 1 \le \alpha  \le 0$ \cite{2}.

For an arbitrary binary sequences $\{ {s_i}\} _{i = 0}^{N - 1}$ of period $N$, let $S(x) = \sum\limits_{i = 0}^{N - 1} {{s_i}} {x^i} \in \mathbb{Z}[x].$ If
\begin{equation}
\frac{{S(2)}}{{{2^N} - 1}} = \frac{{\sum\limits_{i = 0}^{N - 1} {{s_i}{2^i}} }}{{{2^N} - 1}} = \frac{m}{n},0 \le m \le n,\gcd (m,n) = 1,
\end{equation}
then the 2-adic complexity ${\varphi _2}(s)$ of $\{ {s_i}\} _{i = 0}^{N - 1}$, denoting by ${\varphi _2}(s)$, is defined by $\left\lfloor {{{\log }_2}(n + 1)} \right\rfloor $, where $\gcd (m,n)$ is the greatest common divisor of the integers $m$, $n$ and $\left\lfloor {{{\log }_2}n} \right\rfloor $ is the maximal integer no more than ${{{\log }_2}n}$. From formula $(6)$, we know that ${\varphi _2}(s)$ can be calculated by
\begin{equation}
{\varphi _2}(s) = \left\lfloor {{{\log }_2}(\frac{{{2^N} - 1}}{{\gcd ({2^N} - 1,S(2))}})} \right\rfloor .
\end{equation}

\subsection{Gauss period of binary sequences}
\label{sec:2}
Gauss periods were presented by C. F. Gauss in his famous resolution of the problem of constructing regular polygons by straightedge and compass, and have been very useful in studying algebraic structure and in number theory. While Gauss periods can be defined in any finite Galois extension of an arbitrary field, in the following we only consider them over finite field.

Since $a \equiv b(\bmod~{p})$ implies ${2^{2a}} \equiv {2^{2b}}(\bmod~{2^N} - 1)(N = 2p)$, we can define the following mapping
$$f:{\mathbb{Z}_q} \to \mathbb{Z}_{{2^N} - 1}^*,f(a) = {2^a},$$
where $\mathbb{Z}_m^*$ is the group of units in the ring ${\mathbb{Z}_m} = \mathbb{Z}/m\mathbb{Z}(m \ge 2),$ and $f$ is a homomorphism of groups from $({\mathbb{Z}_p}, + ) \to (\mathbb{Z}_{{2^N} - 1}^*, \cdot ),$ clearly, $f$ can also be viewed as an additive character of finite field ${\mathbb{Z}_p} = {\mathbb{F}_p}$ valued in $\mathbb{Z}_{{2^N} - 1}^*.$ The following ${\eta _0}$ and ${\eta _1}$ are called Gauss period of order $2$

$${\eta _0} = \sum\limits_{i \in {D_0}} {{2^i}} \bmod ({2^N} - 1),$$
$${\eta _1} = \sum\limits_{i \in {D_1}} {{2^i}} \bmod ({2^N} - 1).$$

\subsection{Group ring}
\label{sec:2}
Denote $Ho{m_R}(M,N)$ as the group of $\mathbb{R}$-homomorphisms from the $\mathbb{R}$-module $M$ into the $\mathbb{R}$-module $N.$ And $En{d_R}(M) = Ho{m_R}(M,M)$ is the endomorphism ring of the $\mathbb{R}$-module $M.$
Let $\mathbb{R}$ be a ring with unity, $G$ be a group. We consider a free $\mathbb{R}$-module $\mathbb{R}[G]$ with base $\{ g|g \in G\} $, i.e., the correlation of finite formal sums
$$r = \sum\limits_{g \in G} {{r_g} \cdot } g({r_g} \in \mathbb{R},g \in G),$$
with the operations
$$(\sum\limits_{g \in G} {{r_g}g} ) + (\sum\limits_{g \in G} {r_g^{'
}g} ) = \sum\limits_{g \in G} {({r_g} + r_g^{'})g} $$
and
$$r(\sum\limits_{g \in G} {{r_g}g} ) = \sum\limits_{g \in G} {r{r_g}g} $$
which turns into a ring if in it we introduce multiplication by the formula
$$(\sum\limits_{g \in G} {{r_g}g} )(\sum\limits_{g \in G} {r_g^{'}g} ) = \sum\limits_{g \in G} {(\sum\limits_{\begin{array}{*{20}{c}}
{f,h \in G}\\
{fh = g}
\end{array}} {{r_f}{r_h}} )} g.$$

The ring $\mathbb{R}[G]$ is called a group ring of group $G$ over the coefficient ring $\mathbb{R}$. The element $1 \cdot g \in \mathbb{R}[G]$ is called the group base of ring $\mathbb{R}[G]$. Usually instead of $1 \cdot g$ we shall write simply $g$. We denote the unity in a group and in a ring by one and the same symbol $1$, indicating when $1 \in \mathbb{R}$ or $1 \in G$. Sometimes it is convenient for us to describe an element $\tau  \in \mathbb {R}[G]$ in the form
$$\tau  \in \sum {{r_i}{g_l}} ({r_i} \in \mathbb{R},{g_l} \in G).$$
If $R$ is a field, then $\mathbb{R}[G]$ is an $\mathbb{R}$-algebra and in this case $\mathbb{R}[G]$ is called the group algebra of group $G$ over field $\mathbb{R}$.

\section{ Linear complexity over ${\mathbb{F}_q}$ of generalized cyclotomic binary sequences }
\label{sec:1}
\begin{lemma} \cite{19}
Let $D_0^{(p)}$, $D_1^{(p)}$, $D_0^{(2p)}$ and $D_1^{(2p)}$ be the general cyclotomic classes of order $2$ defined previously. Then,

(1) $2 \in D_0^{(p)}$if and only if $p \equiv  \pm 1(\bmod~8)$;

(2) $2 \in D_1^{(p)}$if and only if $p \equiv  \pm 3(\bmod~8)$.
\end{lemma}
\begin{lemma}
Let $D_0^{(p)}$, $D_1^{(p)}$, $D_0^{(2p)}$ and $D_1^{(2p)}$ be the general cyclotomic classes of order $2$ defined previously. Then,
$${2^{ - 1}}(\bmod~{p}) \in \left\{ {\begin{array}{*{20}{c}}
{D_0^{(p)},}&{p \equiv  \pm 1(\bmod~8),}\\
{D_1^{(p)},}&{p \equiv  \pm 3(\bmod~8).}
\end{array}} \right.$$
\end{lemma}

\textbf{\emph{Proof}}
It is easy to be obtained by lemma 1.
\qed

\begin{lemma} \cite{20}
Let $D_0^{(p)}$, $D_1^{(p)}$, $D_0^{(2p)}$and $D_1^{(2p)}$ be the general cyclotomic classes of order $2$ defined previously, $i,j \in \{ 0,1\}. $ Then,

(1) if $a \in D_i^{(p)}$, then $aD_j^{(p)}(\bmod~{p}) = D_{i + j(\bmod~2)}^{(p)};$

(2) $D_i^{(2p)}(\bmod~{p}) = D_i^{(p)}.$
\end{lemma}

\begin{lemma}
Let $D_0^{(p)}$, $D_1^{(p)}$, $D_0^{(2p)}$ and $D_1^{(2p)}$ be the general cyclotomic classes of order $2$ defined previously, $\beta$ be a $2p^{th}$ root of unity. Then,
$${\beta ^{pk}} \equiv \left\{ {\begin{array}{*{20}{c}}
{ - 1(\bmod~{r}),}&{k \in D_0^{(2p)} \cup D_1^{(2p)} \cup \{ p\} ,}\\
{1(\bmod~{r}),}&{k \in 2D_0^{(p)} \cup 2D_1^{(p)} \cup \{ 0\} .}
\end{array}} \right.$$
\end{lemma}

\textbf{\emph{Proof}}
Since $\beta$ be a $2p^{th}$ root of unity, ${\beta ^p} = - 1$. Therefore, if $k$ is an even, ${\beta ^{pk}} = 1$; if $k$ be an odd, ${\beta ^{pk}} = -1.$
\qed

\begin{lemma}
Let
\begin{align*}
{M_p} &  = 2 \cdot ({2^{ - 1}}(\bmod~{p}))(\bmod~{2p}),\\
{M_2} &  = p \cdot ({p^{ - 1}}(\bmod~2))(\bmod~{2p}),\\
\mathbb{Z}_2^* & = \{ 1\},
\end{align*}
then,
$$\sum\limits_{t \in D_1^{(2p)}} {{\beta ^{kt}}}  \equiv {\beta ^{pk}}\sum\limits_{j \in D_1^{(p)}} {{\beta ^{2({2^{ - 1}}\bmod p)kj}}} (\bmod~{r}),$$
where $k \in D_0^{(2p)} \cup D_1^{(2p)} \cup 2D_0^{(p)} \cup 2D_1^{(p)}.$
\end{lemma}

\textbf{\emph{Proof}}
Let $\mathbb{Z}_2^*$ denote the set of all invetible elements in $\mathbb{Z}_2$. It is obvious that $\mathbb{Z}_2^* = \{ 1\}$. By the Chinese remainder theorem, we have
$$D_1^{(2p)} \cong \mathbb{Z}_2^* \times D_1^{(p)},$$
where $t \in D_1^{(2p)}$ and $j \equiv t(\bmod~{p}).$ From the Chinese remainder theorem, $t \equiv 1 \cdot {M_2} + j \cdot {M_p}(\bmod~{2p}) \equiv 1 \cdot p + j \cdot 2 \cdot ({2^{ - 1}}\bmod p)(\bmod~{2p}).$
Therefore,
    \begin{align*}
	    \sum\limits_{t \in D_1^{(2p)}} {{\beta ^{kt}}} & \equiv \sum\limits_{j \in D_1^{(p)}} {{\beta ^{k(p + j \cdot 2({2^{ - 1}}\bmod p))}}} \\
	& \equiv {\beta ^{pk}}\sum\limits_{j \in D_1^{(p)}} {{\beta ^{2({2^{ - 1}}\bmod p)kj}}} (\bmod~{r}).
	\end{align*}
\qed

\begin{lemma} \cite{21}
Let $D_0^{(p)}$, $D_1^{(p)}$, $D_0^{(2p)}$ and $D_1^{(2p)}$ be the general cyclotomic classes of order $2$ defined previously, r be an odd prime, $r \ge 5$ and $r \ne p$, where $p$ is an odd prime, $\beta$ be a $2p^{th}$ root of unity. Denote Gauss period
$${\eta _0} = \sum\limits_{i \in D_0^{(p)}} {{\beta ^{2i}}}, {\eta _1} = \sum\limits_{i \in D_1^{(p)}} {{\beta ^{2i}}}.$$
Then, ${\eta _0} \in {\mathbb{F}_r}$ and ${\eta _1} \in {\mathbb{F}_r}$ if and only if $r \in D_0^{(p)}.$
\end{lemma}

\begin{lemma}
Let $D_0^{(p)}$, $D_1^{(p)}$, $D_0^{(2p)}$ and $D_1^{(2p)}$ be the general cyclotomic classes of order $2$ defined previously, r be an odd prime, $r \ge 5$ and $r \ne p$, where $p$ is an odd prime, $\beta$ be a $2p^{th}$ root of unity, ${\eta _0}$ and ${\eta _1}$ is defined the same as lemma 6,

(1) ${\eta _0} + {\eta _1} =  - 1.$

(2) \cite{22} if $p \equiv 1(\bmod~4)$, then ${\eta _0}(1 + {\eta _0}) \equiv \frac{{p - 1}}{4}(\bmod~{r});$

if $p \equiv 3(\bmod~4)$, then ${\eta _1}(1 + {\eta _1}) \equiv  - \frac{{p + 1}}{4}(\bmod~{r}).$
\end{lemma}
\textbf{\emph{Proof}}
(1) From
$${\beta ^0} + {\beta ^p} + \sum\limits_{t \in 2D_0^{(p)}} {{\beta ^t} + } \sum\limits_{t \in 2D_1^{(p)}} {{\beta ^t} + } \sum\limits_{t \in D_0^{(2p)}} {{\beta ^t} + } \sum\limits_{t \in D_1^{(2p)}} {{\beta ^t}}  = 0,$$
we have
$$\sum\limits_{t \in 2D_0^{(p)}} {{\beta ^{kt}}}  + \sum\limits_{t \in 2D_1^{(p)}} {{\beta ^{kt}}}  =  - 1,$$
where $k \in D_0^{(2p)} \cup D_1^{(2p)} \cup 2D_0^{(p)} \cup 2D_1^{(p)}.$ That is,
$${\eta _0} + {\eta _1} =  - 1.$$
\qed

According to the definition of generating polynomial of sequences $\{ {s_i}\} _{i = 0}^{N - 1}$ previously,
\begin{equation}
S(x) = \sum\limits_{i \in {C_1}} {{x^i}}  = 1 + \sum\limits_{i \in D_1^{(2p)}} {{x^i}}  + \sum\limits_{i \in 2D_1^{(p)}} {{x^i}} .
\end{equation}

\begin{lemma}
Let $D_0^{(p)}$, $D_1^{(p)}$, $D_0^{(2p)}$ and $D_1^{(2p)}$ be the general cyclotomic classes of order $2$ defined previously, r be an odd prime, $r \ge 5$ and $r \ne p$, where $p$ is an odd prime. Let $\beta$ be a $2p^{th}$ root of unity, ${\eta _0}$ and ${\eta _1}$ be defined the same as lemma 6. Then,

(1) when $p \equiv  \pm 1(\bmod~8),$
$$S({\beta ^k}) = \left\{ {\begin{array}{*{20}{l}}
{p(\bmod~{r}),}&{k = 0,}\\
{1(\bmod~{r}),}&{k \in D_0^{(2p)} \cup D_1^{(2p)} \cup \left\{ p \right\},}\\
{1 + 2{\eta _1}(\bmod~{r}),}&{k \in 2D_0^{(p)},}\\
{1 + 2{\eta _0}(\bmod~{r}),}&{k \in 2D_1^{(p)};}
\end{array}} \right.$$

(2) when $p \equiv  \pm 3(\bmod~8),$
$$S({\beta ^k}) = \left\{ {\begin{array}{*{20}{l}}
{p(\bmod~{r}),}&{k = 0,}\\
{1(\bmod~{r}),}&{k = p,}\\
{ - 2{\eta _0}(\bmod~{r}),}&{k \in D_0^{(2p)},}\\
{ - 2{\eta _1}(\bmod~{r}),}&{k \in D_1^{(2p)},}\\
{0(\bmod~{r}),}&{k \in 2D_0^{(p)} \cup 2D_1^{(p)}.}
\end{array}} \right.$$

\end{lemma}
\textbf{\emph{Proof}}
Let $\alpha  = {\beta ^2}. $ Since $\beta$ is a $2p^{th}$ root of unity, we known that $\alpha $ is a $p^{th}$ root of unity.
(i)For $k = 0$, from formula(8),
    \begin{align*}
	    S({\beta ^k}) & = \sum\limits_{i \in {C_1}} {{\beta ^{ki}}}  = 1 + \sum\limits_{i \in D_1^{(2p)}} {{\beta ^{ki}}}  + \sum\limits_{i \in 2D_1^{(p)}} {{\beta ^{ki}}} \\
	& = 1 + \sum\limits_{i \in D_1^{(2p)}} 1  + \sum\limits_{i \in 2D_1^{(p)}} 1 \\
	& \equiv p(\bmod~{r}).
	\end{align*}
(ii)For $k = p$, from formula(8) and lemma 4, we have
    \begin{align*}
	    S({\beta ^k}) & = 1 + \sum\limits_{i \in D_1^{(2p)}} {{\beta ^{ki}}}  + \sum\limits_{i \in 2D_1^{(p)}} {{\beta ^{ki}}} \\
	& = 1 + \sum\limits_{i \in D_1^{(2p)}} {( - 1)}  + \sum\limits_{i \in 2D_1^{(p)}} 1 \\
	& \equiv 1(\bmod~{r}).
	\end{align*}
(iii) For $k \in D_0^{(2p)} \cup D_1^{(2p)} \cup 2D_0^{(p)} \cup 2D_1^{(p)}$, denote $2_p^{ - 1} = {2^{ - 1}}(\bmod~{p})$,

1) when $p \equiv  \pm 1(\bmod~8)$, from lemma 2 known that $2_p^{ - 1} \in D_0^{(p)}.$

(a) When $k \in D_0^{(2p)} \cup D_1^{(2p)},$  by lemma 1, 3, 4, 5, we obtain
    \begin{align*}
	    S({\beta ^k}) & = 1 + \sum\limits_{t \in D_1^{(2p)}} {{\beta ^{kt}}}  + \sum\limits_{t \in 2D_1^{(p)}} {{\beta ^{kt}}} \\
	& = 1 + {\beta ^{kp}}\sum\limits_{j \in D_1^{(p)}} {{\beta ^{2({2^{ - 1}}\bmod p)kj}}}  + \sum\limits_{t \in 2D_1^{(p)}} {{\beta ^{kt}}} \\
	& = 1 - \sum\limits_{j \in D_1^{(p)}} {{{({\beta ^k})^{2(2_p^{ - 1})j}}}}  + \sum\limits_{t \in 2D_1^{(p)}} {{\beta ^{kt}}} \\
	& = 1 - \sum\limits_{t \in 2_p^{ - 1}D_1^{(p)}} {{{({\beta ^k})^{2t}}}}  + \sum\limits_{t \in 2D_1^{(p)}} {{\beta ^{kt}}} \\
	& = 1 - \sum\limits_{t \in D_1^{(p)}} {{{({\beta ^k})^{2t}}}}  + \sum\limits_{t \in 2D_1^{(p)}} {{\beta ^{kt}}} \\
	& = 1 - \sum\limits_{t \in D_1^{(p)}} {{{({\beta ^2})^{kt}}}}  + \sum\limits_{t \in D_1^{(p)}} {{{({\beta ^2})^{kt}}}} \\
	& = 1 - \sum\limits_{t \in D_1^{(p)}} {{{(\alpha )}^{kt}}}  + \sum\limits_{t \in D_1^{(p)}} {{{(\alpha )}^{kt}}} \\
	& \equiv 1(\bmod~{r}).
	\end{align*}
(b) When $k \in 2D_0^{(p)},$  by lemma 1, 3, 4, we obtain
    \begin{align*}
	    S({\beta ^k}) & = 1 + \sum\limits_{t \in D_1^{(2p)}} {{\beta ^{kt}}}  + \sum\limits_{t \in 2D_1^{(p)}} {{\beta ^{kt}}} \\
	& = 1 + {\beta ^{kp}}\sum\limits_{j \in D_1^{(p)}} {{\beta ^{2({2^{ - 1}}\bmod p)kj}}}  + \sum\limits_{t \in 2D_1^{(p)}} {{\beta ^{kt}}} \\
	& = 1 + \sum\limits_{j \in D_1^{(p)}} {{{({\beta ^k})^{2(2_p^{ - 1})j}}}}  + \sum\limits_{t \in 2D_1^{(p)}} {{\beta ^{kt}}} \\
	& = 1 + \sum\limits_{t \in 2_p^{ - 1}D_1^{(p)}} {{{({\beta ^k})^{2t}}}}  + \sum\limits_{t \in 2D_1^{(p)}} {{\beta ^{kt}}} \\
	& = 1 + \sum\limits_{t \in D_1^{(p)}} {{{({\beta ^k})^{2t}}}}  + \sum\limits_{t \in D_1^{(p)}} {{{({\beta ^k})^{2t}}}} {\rm{ }}\\
	& = 1 + 2\sum\limits_{t \in D_1^{(p)}} {{{({\beta ^k})^{2t}}}} {\rm{ }}\\
	& = 1 + 2\sum\limits_{t \in D_1^{(p)}} {{{(\alpha )}^{kt}}}.
	\end{align*}
In this case, there exists an $n \in D_0^{(p)}$ such that $k = 2n$. By lemma 1, 3, $2D_0^{(p)}(\bmod~{p}) = D_0^{(p)},2D_1^{(p)}(\bmod~{p}) = D_1^{(p)}, nD_0^{(p)}(\bmod~{p}) = D_0^{(p)},nD_1^{(p)}(\bmod~{p}) = D_1^{(p)}.$  Therefore,
    \begin{align*}
	    S({\beta ^k}) & = 1 + 2\sum\limits_{t \in D_1^{(p)}} {{{(\alpha )}^{2nt}}} \\
	& = 1 + 2\sum\limits_{t \in 2D_1^{(p)}} {{{(\alpha )}^{nt}}} \\
	& = 1 + 2\sum\limits_{t \in D_1^{(p)}} {{{(\alpha )}^{nt}}} \\
	& = 1 + 2\sum\limits_{t \in nD_1^{(p)}} {{{(\alpha )}^t}} \\
	& = 1 + 2\sum\limits_{t \in D_1^{(p)}} {{{(\alpha )}^t}} \\
	& \equiv 1 + 2{\eta _1}(\bmod~{r}).
	\end{align*}
(c) When $k \in 2D_1^{(p)},$ using the same method as that in (b), we obtain
$$S({\beta ^k}) = 1 + 2\sum\limits_{t \in D_1^{(p)}} {{{(\alpha )}^{kt}}}. $$
In this case, there exists an $n \in D_1^{(p)}$ such that $k = 2n$. By lemma 1, 3, $2D_0^{(p)}(\bmod~{p}) = D_0^{(p)},2D_1^{(p)}(\bmod~{p}) = D_1^{(p)}, nD_0^{(p)}(\bmod~{p}) = D_1^{(p)},nD_1^{(p)}(\bmod~{p}) = D_0^{(p)}.$  Therefore,
    \begin{align*}
	    S({\beta ^k}) & = 1 + 2\sum\limits_{t \in D_1^{(p)}} {{{(\alpha )}^{2nt}}} \\
	& = 1 + 2\sum\limits_{t \in 2D_1^{(p)}} {{{(\alpha )}^{nt}}} \\
	& = 1 + 2\sum\limits_{t \in D_1^{(p)}} {{{(\alpha )}^{nt}}} \\
	& = 1 + 2\sum\limits_{t \in nD_1^{(p)}} {{{(\alpha )}^t}} \\
	& = 1 + 2\sum\limits_{t \in D_0^{(p)}} {{{(\alpha )}^t}} \\
	& \equiv 1 + 2{\eta _0}(\bmod~{r}){\rm{ }}.
	\end{align*}
2) When $p \equiv  \pm 3(\bmod~8)$, from lemma 1 we known that $2_p^{ - 1} \in D_1^{(p)}.$

(a) When $k \in D_0^{(2p)},$  by lemma 1, 3, 4, 5, 7, we obtain
    \begin{align*}
	    S({\beta ^k}) & = 1 - \sum\limits_{t \in 2_p^{ - 1}D_1^{(p)}} {{{({\beta ^k})^{2t}}}}  + \sum\limits_{t \in 2D_1^{(p)}} {{\beta ^{kt}}} \\
	& = 1 - \sum\limits_{t \in D_0^{(p)}} {{{({\beta ^k})^{2t}}}}  + \sum\limits_{t \in 2D_1^{(p)}} {{\beta ^{kt}}} \\
	& = 1 - \sum\limits_{t \in D_0^{(p)}} {{\alpha ^{kt}}}  + \sum\limits_{t \in D_1^{(p)}} {{\alpha ^{kt}}} {\rm{ }}\\
	& = 1 - \sum\limits_{t \in D_0^{(p)}} {{\alpha ^t}}  + \sum\limits_{t \in D_1^{(p)}} {{\alpha ^t}} {\rm{ }}\\
	& = 1 - {\eta _0} + {\eta _1}{\rm{ }}\\
	& \equiv  - 2{\eta _0}(\bmod~{r}){\rm{  }}.
	\end{align*}
(b) When $k \in D_1^{(2p)},$  by lemma 1, 3, 4, 5, 7, we obtain
    \begin{align*}
	    S({\beta ^k}) & = 1 - \sum\limits_{t \in D_0^{(p)}} {{\alpha ^{kt}}}  + \sum\limits_{t \in D_1^{(p)}} {{\alpha ^{kt}}} \\
	& = 1 - \sum\limits_{t \in D_1^{(p)}} {{\alpha ^t}}  + \sum\limits_{t \in D_0^{(p)}} {{\alpha ^t}} {\rm{ }}\\
	& = 1 - {\eta _1} + {\eta _0}{\rm{ }}\\
	& \equiv  - 2{\eta _1}(\bmod~{r}){\rm{  }}.
	\end{align*}
(c) When $k \in 2D_0^{(p)},$  by lemma 1, 3, 4, 5, 7, we obtain
    \begin{align*}
	    S({\beta ^k}) & = 1 + \sum\limits_{t \in 2_p^{ - 1}D_1^{(p)}} {{{({\beta ^k})^{2t}}}}  + \sum\limits_{t \in 2D_1^{(p)}} {{\beta ^{kt}}} \\
	& = 1 + \sum\limits_{t \in D_0^{(p)}} {{{({\beta ^k})^{2t}}}}  + \sum\limits_{t \in D_1^{(p)}} {{\beta ^{2kt}}} \\
	& = 1 + \sum\limits_{t \in D_0^{(p)}} {{\alpha ^{kt}}}  + \sum\limits_{t \in D_1^{(p)}} {{\alpha ^{kt}}} {\rm{ }}.
	\end{align*}
In this case, there exists an $n \in D_0^{(p)}$ such that $k = 2n$. By lemma 1, 3, $2D_0^{(p)}(\bmod~{p}) = D_1^{(p)},2D_1^{(p)}(\bmod~{p}) = D_0^{(p)}, nD_0^{(p)}(\bmod~{p}) = D_0^{(p)},nD_1^{(p)}(\bmod~{p}) = D_1^{(p)}.$  Therefore,
	\begin{align*}
	    S({\beta ^k}) & = 1 + \sum\limits_{t \in D_0^{(p)}} {{{(\alpha )}^{2nt}}}  + \sum\limits_{t \in D_1^{(p)}} {{{(\alpha )}^{2nt}}} \\
	& = 1 + \sum\limits_{t \in 2D_0^{(p)}} {{{(\alpha )}^{nt}} + \sum\limits_{t \in 2D_1^{(p)}} {{{(\alpha )}^{nt}}} } \\
	& = 1 + \sum\limits_{t \in D_1^{(p)}} {{{(\alpha )}^{nt}} + \sum\limits_{t \in D_0^{(p)}} {{{(\alpha )}^{nt}}} } \\
	& = 1 + \sum\limits_{t \in D_1^{(p)}} {{\alpha ^t} + \sum\limits_{t \in D_0^{(p)}} {{\alpha ^t}} } \\
	& = 1 + {\eta _1} + {\eta _0}\\
	& \equiv 0(\bmod~{r}){\rm{     }}.
	\end{align*}
(d) When $k \in 2D_1^{(p)},$  using the similar method as that in (c), we obtain

$$S({\beta ^k}) \equiv 1 + \sum\limits_{t \in D_0^{(p)}} {{\alpha ^{kt}}}  + \sum\limits_{t \in D_1^{(p)}} {{\alpha ^{kt}}}. $$
In this case, there exists an $n \in D_1^{(p)}$ such that $k = 2n$. By lemma 1, 3, $2D_0^{(p)}(\bmod~{p}) = D_1^{(p)},2D_1^{(p)}(\bmod~{p}) = D_0^{(p)}, nD_0^{(p)}(\bmod~{p}) = D_1^{(p)},nD_1^{(p)}(\bmod~{p}) = D_0^{(p)}.$  Therefore,
	\begin{align*}
	    S({\beta ^k}) & =1 + \sum\limits_{t \in D_1^{(p)}} {{{(\alpha )}^{nt}} + \sum\limits_{t \in D_0^{(p)}} {{{(\alpha )}^{nt}}} } \\
	& = 1 + \sum\limits_{t \in nD_1^{(p)}} {{\alpha ^t} + \sum\limits_{t \in nD_0^{(p)}} {{\alpha ^t}} } \\
	& = 1 + \sum\limits_{t \in D_0^{(p)}} {{\alpha ^t} + \sum\limits_{t \in D_1^{(p)}} {{\alpha ^t}} } \\
	& = 1 + {\eta _0} + {\eta _1}\\
	& \equiv 0(\bmod~{r}){\rm{     }}.
	\end{align*}
\qed

\begin{theorem}
Let $r$ be an odd prime such that $r \ge 5$ and $r \ne p$, $m$ be the order of $r$ module $p$, and $\beta $ be a $2{p^{th}}$ root of unity. In addition, let $\left( {\frac{r}{p}} \right)$ denote the Legendre symbol of $r$ modulo $p$. Then, the linear complexity over ${\mathbb{F}_q}$ of a class of generalized cyclotomic binary sequences which were defined previously is

$$LC(s) = \left\{ {\begin{array}{*{20}{c}}
{2p,}&{p \equiv  \pm 1(\bmod~8),}\\
{p + 1,}&{p \equiv  \pm 3(\bmod~8).}
\end{array}} \right.$$
\end{theorem}

\textbf{\emph{Proof}}
Denote $S(x)$ as the generating polynomial of $\left\{ {{s_i}} \right\}_{i = 0}^{N - 1}$, in order to determine the linear complexity of the generalized cyclotomic binary sequences of period $2p$ defined in formula(1), we must known the degree of $\gcd ({x^{2p}} - 1,S(x))$. Since $\beta $ is a $2{p^{th}}$ root of unity, ${\beta ^k}$ is a zero of polynomial ${x^{2p}} - 1 \in {\mathbb{F}_q} (q = {r^m})$ for each $k \in {\mathbb{Z}_{2p}}$. If the same ${\beta ^k}$ is also the zero of $S(x)$, it means $x - {\beta ^k}$ is a common factor of ${x^{2p}} - 1$ and $S(x)$. Therefore, we can determine the degree of $\gcd ({x^{2p}} - 1,S(x))$ by computing $|\{ k:S({\beta ^k}) = 0\} |$.

(1) When $p \equiv  \pm 1(\bmod~8)$, we have

  (a) If $\left( {\frac{r}{p}} \right) = 1$, then, from lemma 6, ${\eta _0} \in {\mathbb{F}_r}$ and $ {\eta _1} \in {\mathbb{F}_r}$. According to lemma 8, we can obtain $S({\beta ^k}) \equiv 1 + 2{\eta _1}(\bmod~{r})$ where $k \in 2D_0^{(p)}$; $S({\beta ^k}) \equiv 1 + 2{\eta _0}(\bmod~{r})$ where $k \in 2D_1^{(p)}$.

According to lemma 7 and by resolving the following two congruent systems, we can determine the prime $p$ by which ${\beta ^k}$, the zeros of $S(x)$, is obtained.

For $k \in 2D_0^{(p)}$,

\begin{equation}
\left\{ {\begin{array}{*{20}{c}}
{1 + 2{\eta _1} \equiv 0(\bmod~{r}),}\\
{{\eta _0} + {\eta _1} \equiv  - 1(\bmod~{r}),}\\
{{\eta _0}(1 + {\eta _0}) \equiv \frac{{p - 1}}{4}(\bmod~{r}).}
\end{array}} \right.
\end{equation}

For $k \in 2D_1^{(p)}$,
\begin{equation}
\left\{ {\begin{array}{*{20}{c}}
{1 + 2{\eta _0} \equiv 0(\bmod~{r}),}\\
{{\eta _0} + {\eta _1} \equiv  - 1(\bmod~{r}),}\\
{{\eta _0}(1 + {\eta _0}) \equiv \frac{{p - 1}}{4}(\bmod~{r}).}
\end{array}} \right.
\end{equation}

The solution by resolving the formula(9) is
\begin{equation}
\left\{ {\begin{array}{*{20}{c}}
{{\eta _1} \equiv  - {2^{ - 1}}(\bmod~{r}),}\\
{{\eta _0} \equiv  - {2^{ - 1}}(\bmod~{r}),}\\
{p \equiv 0(\bmod~{r}).}
\end{array}} \right.
\end{equation}

The solution by resolving the formula(10) is
\begin{equation}
\left\{ {\begin{array}{*{20}{c}}
{{\eta _1} \equiv  - {2^{ - 1}}(\bmod~{r}),}\\
{{\eta _0} \equiv  - {2^{ - 1}}(\bmod~{r}),}\\
{p \equiv 0(\bmod~{r}).}
\end{array}} \right.
\end{equation}

(b) If $\left( {\frac{r}{p}} \right) = - 1$, then, since $r \ne p$, by lemma 8, the minimal polynomial of sequences $\{ {s_i}\} _{i = 0}^{N - 1}$ is
$$m(x) = {x^{2p}} - 1,$$
and the linear complexity of sequences $\{ {s_i}\} _{i = 0}^{N - 1}$ is
$$LC(s) = \deg (m(x)) = 2p.$$

(2) When $p \equiv  \pm 3(\bmod~8)$, the minimal polynomial of sequences $\{ {s_i}\} _{i = 0}^{N - 1}$ is
$$m(x) = \frac{{{x^{2p}} - 1}}{{\prod\limits_{k \in 2D_0^{(p)} \cup 2D_1^{(p)}} {(x - {\beta ^k})} }},$$
and the linear complexity of sequences $\left\{ {{s_i}} \right\}_{i = 0}^{N - 1}$ is
$$LC(s) = 2p - \frac{{p - 1}}{2} - \frac{{p - 1}}{2} = p + 1,$$
which is larger than the half of the class of generalized cyclotomic sequences of period $2p$ defined previously.
\qed
\textbf{\emph{Example} 1}
Let $ p = 5,g = 3,r = 7. $ Then,
$$ D_0^{(2p)} = \{ 1,9\} , D_1^{(2p)} = \{ 3,7\} , 2D_0^{(p)} = \{ 2,8\} ,  2D_1^{(p)} = \{ 4,6\} .$$
The corresponding generalized cyclotomic binary sequence of period $10$ is as follows:
$$s = 1001101100.$$

By using Magma, the linear complexity and the the minimal polynomial of the above sequence are respectively
$$ LC(s) = 6 = p + 1,    m(x) = \frac{{{x^{10}} - 1}}{{(x - {\beta ^2})(x - {\beta ^4})(x - {\beta ^6})(x - {\beta ^8})}} . $$

\textbf{\emph{Example} 2}
Let $ p = 13,g = 7,r = 5. $ Then,
$$ D_0^{(2p)} = \left\{ {1,3,9,17,23,25} \right\} , D_1^{(2p)} = \left\{ {5,7,11,15,19,21} \right\} ,$$
$$ 2D_0^{(p)} = \left\{ {2,6,8,18,20,24} \right\} , 2D_1^{(p)} = \{ 4,10,12,14,16,22\}  .$$
The corresponding generalized cyclotomic binary sequence of period $ 26 $ is as follows:
$$ s = 10001101001110111001011000. $$

By using Magma, the linear complexity and the the minimal polynomial of the above sequence are respectively
$$ LC(s) = 14 = p + 1 ,    m(x) = \frac{{{x^{26}} - 1}}{{\prod\limits_{k \in 2D_0^{(p)} \cup 2D_1^{(p)}} {(x - {\beta ^k})} }} .$$

\textbf{\emph{Example} 3}
Let $ p = 17,g = 3,r = 5. $ Then,
$$ D_0^{(2p)} = \left\{ {1,9,13,15,19,21,25,33} \right\} , D_1^{(2p)} = \left\{ {\;3,5,7,11,23,27,29,31} \right\} ,$$
$$ 2D_0^{(p)} = \left\{ {2,4,8,16,18,26,30,32} \right\} , 2D_1^{(p)} = \{ \;6,10,12,14,20,22,24,28\}  .$$
The corresponding generalized cyclotomic binary sequence of period $ 34 $ is as follows:
$$ s = 1001011100111010000010111001110100. $$

By using Magma, the linear complexity and the the minimal polynomial of the above sequence are respectively
$$ LC(s) = 34 = 2p ,    m(x) = {x^{34}} - 1 . $$

\textbf{\emph{Example} 4}
Let $ p = 19,g = 3,r = 13. $ Then,
$$ D_0^{(2p)} = \left\{ {1,5,7,9,11,17,23,25,35} \right\} , D_1^{(2p)} = \left\{ {\;3,5,7,11,23,27,29,31} \right\} ,$$
$$ 2D_0^{(p)} = \left\{ {2,8,10,12,14,18,22,32,34} \right\} , 2D_1^{(p)} = \{ \;4,6,16,20,24,26,{\rm{2}}8,30,36\}  .$$
The corresponding generalized cyclotomic binary sequence of period $ 38 $ is as follows:
$$ s = 10011010000001011000110010111111010011. $$

By using Magma, the linear complexity and the the minimal polynomial of the above sequence are respectively
$$ LC(s) = 20 = p + 1 ,    m(x) = \frac{{{x^{38}} - 1}}{{\prod\limits_{k \in 2D_0^{(p)} \cup 2D_1^{(p)}} {(x - {\beta ^k})} }} . $$

\textbf{\emph{Example} 5}
Let $ p = 23,g = 7,r = 13. $ Then,
$$ D_0^{(2p)} = \left\{ {1,3,9,13,25,27,29,31,35,39,41} \right\} , D_1^{(2p)} = \left\{ {\;5,7,11,15,17,19,21,33,37,43,45} \right\}, $$
$$ 2D_0^{(p)} = \left\{ {2,4,6,8,12,16,18,24,26,32,36} \right\} , 2D_1^{(p)} = \left\{ {10,14,20,22,28,30,34,38,40,42,44} \right\}  .$$
The corresponding generalized cyclotomic binary sequence of period $ 46 $ is as follows:
$$ \begin{array}{l}
s = 1000010100110011010111100000101001100110101111.
\end{array} $$

By using Magma, the linear complexity and the the minimal polynomial of the above sequence are respectively
$$ LC(s) = 46 = 2p ,    m(x) = {x^{46}} - 1 . $$

\textbf{\emph{Example} 6}
Let $ p = 113,g = 7,r = 13. $ Then,
$$ \begin{array}{l}
D_0^{(2p)} = \{ 1,7,9,11,13,15,25,31,41,49,51,53,57,61,63,69,77,81,83,85,87,91,95,97,99,\\
105,109,111,115,117,121,127,129,131,135,139,141,143,145,149,157,163,165,169,173,\\
175,177,185,195,201,211,213,215,217,219,225\} ,
\end{array} $$
$$ \begin{array}{l}
D_1^{(2p)} = \{ 3,5,17,19,21,23,27,29,33,35,37,39,43,45,47,55,59,65,67,71,73,75,79,89,93,\\
101,103,107,119,123,125,133,137,147,151,153,155,159,161,167,171,179,181,183,187,\\
189,191,193,197,199,203,205,207,209,221,223\} ,
\end{array} $$
$$ \begin{array}{l}
2D_0^{(p)} = \{ 2,4,8,14,16,18,22,26,28,30,32,36,44,50,52,56,60,62,64,72,82,88,98,100,102,\\
104,106,112,114,120,122,124,126,128,138,144,154,162,164,166,170,174,176,182,190,\\
194,196,198,200,204,208,210,212,218,222,224\} ,
\end{array} $$
$$ \begin{array}{l}
2D_1^{(p)} = \{ 6,10,12,20,24,34,38,40,42,46,48,54,58,66,68,70,74,76,78,80,84,86,90,92,94,\\
96,108,110,116,118,130,132,134,136,140,142,146,148,150,152,156,158,160,168,172,\\
178,180,184,186,188,192,202,206,214,216,220\}.
\end{array}  $$
The corresponding generalized cyclotomic binary sequence of period $ 226 $ is as follows:
$$ \begin{array}{l}
\begin{array}{l}
s = 1001011000101000010111011001010001110111101101111000001100110000011110110\\
11110111000101001101110100001010001101000001011000101000010111011001010001110\\
1111011011110000011001100000111101101111011100010100110111010000101000110100 .
\end{array}
\end{array} $$

By using Magma, the linear complexity and the the minimal polynomial of the above sequence are respectively
$$ LC(s) = 226 = 2p ,   m(x) = {x^{226}} - 1 . $$

\section{ 2-adic complexity of generalized cyclotomic binary sequences }
\label{sec:1}
Let $p$ be an odd prime, $N = 2p$. The fact $a \equiv b(\bmod~{p})$ implies ${2^{2a}} \equiv {2^{2b}}(\bmod {2^{N}} - 1)$. Then, we can define an element ${G_p}$ in group ring ${\mathbb{Z}_{{2^N} - 1}}$:
\begin{equation*}
{G_p} = \sum\limits_{a \in \mathbb{Z}_p^*} {\left( {\frac{a}{p}} \right)} {2^{2a}} = \sum\limits_{a = 1}^{p - 1} {\left( {\frac{a}{p}} \right){2^{2a}}} (\bmod~{2^N} - 1).
\end{equation*}
In order to compute ${\varphi _2}(s)$, we first prove the following lemmas.

\begin{lemma}
 Let $s = \{ {s_0},{s_1},{s_2}, \cdots, {s_{N - 1}}\} $ be the binary sequences over ${\mathbb{Z}_2}$ with period $N = 2p$ defined by formula$(1)$. Then,

(1) $S(2) \equiv ( - \frac{{{2^p} + 1}}{2} + 1) + (\frac{{{2^p} + 1}}{2})(\frac{{{2^{2p}} - 1}}{3}) - \frac{1}{2}\left( {\left( {\frac{2}{p}} \right){2^p} + 1} \right){G_p}(\bmod~{2^N} - 1)$;

(2) $G_p^2 \equiv \left( {\frac{{ - 1}}{p}} \right)(p - \frac{{{2^N} - 1}}{3})(\bmod~{2^N} - 1)$.

\end{lemma}

\textbf{\emph{Proof}}
(1) According to the Chinese remainder theorem, we have isomorphism of rings
\begin{equation*}
\varphi :{\mathbb{Z}_{2p}} \cong {\mathbb{Z}_p} \oplus {\mathbb{Z}_2},
\end{equation*}
where $\varphi (x(\bmod~{2p})) = (x(\bmod~{p}),x(\bmod~2))$. Solving the following congruent system of equations
$$\left\{ {\begin{array}{*{20}{c}}
{x \equiv A(\bmod~{p})},\\
{x \equiv B(\bmod~2)},
\end{array}} \right.$$
we obtain for any element $(A,B) \in {\mathbb{Z}_p} \oplus {\mathbb{Z}_2}, i.e., 0 \le A \le p - 1,B \in \{ 0,1\}, {\varphi ^{ - 1}}(A,B) = A(p + 1) + pB \in {\mathbb{Z}_{2p}}$. Then,
\begin{equation*}
\sum\limits_{i \in D_1^{(2p)}} {{2^i}}  \equiv \sum\limits_{\begin{array}{*{20}{c}}
{A = 1}\\
{\left( {\frac{A}{p}} \right) =  - 1}
\end{array}}^{p - 1} {{2^{A(p + 1) + p}}}  \equiv \sum\limits_{\begin{array}{*{20}{c}}
{A = 1}\\
{\left( {\frac{{2A}}{p}} \right) =  - 1}
\end{array}}^{p - 1} {{2^{2A + p}}} (\bmod~{2^{2p}} - 1),
\end{equation*}
where $i = A(p + 1) + p$.
From formula$(2)$, we have
	
\begin{align*}
	    S(2) & = \sum\limits_{i \in D_1^{(2p)}} {{2^i}}  + \sum\limits_{i \in 2D_1^{(p)}} {{2^i}}  + 1\\
	& \equiv \sum\limits_{\begin{array}{*{20}{c}}
{a = 1}\\
{\left( {\frac{{2a}}{p}} \right) =  - 1}
\end{array}}^{p - 1} {{2^{2a + p}}}  + \sum\limits_{\begin{array}{*{20}{c}}
{a \in D_1^{(p)}}\\
{\left( {\frac{a}{p}} \right) =  - 1}
\end{array}} {{2^{2a}}}  + 1(\bmod~{2^N} - 1)\\
	& \equiv {2^p} \cdot \sum\limits_{\begin{array}{*{20}{c}}
{a = 1}\\
{\left( {\frac{{a}}{p}} \right) =  - \left( {\frac{2}{p}} \right)}
\end{array}}^{p - 1} {{2^{2a}}}  + \sum\limits_{\begin{array}{*{20}{c}}
{a \in D_1^{(p)}}\\
{\left( {\frac{a}{p}} \right) =  - 1}
\end{array}} {{2^{2a}}}  + 1(\bmod~{2^N} - 1)\\
	& \equiv {2^p} \cdot \frac{1}{2}\sum\limits_{a = 1}^{p - 1} {\left( {1 - \left( {\frac{2}{p}} \right)\left( {\frac{a}{p}} \right)} \right){2^{2a}}}  + \frac{1}{2}\sum\limits_{a = 1}^{p - 1} {\left( {1 - \left( {\frac{a}{p}} \right)} \right){2^{2a}}}  + 1(\bmod~{2^N} - 1)\\
& \equiv (\frac{{{2^p}}}{2} + \frac{1}{2})\sum\limits_{a = 1}^{p - 1} {{2^{2a}} - (\frac{{{2^p}}}{2}\left( {\frac{2}{p}} \right) + \frac{1}{2})\sum\limits_{a = 1}^{p - 1} {\left( {\frac{a}{p}} \right)} {2^{2a}}}  + 1(\bmod~{2^N} - 1)\\
	& \equiv (\frac{{{2^p} + 1}}{2})\sum\limits_{a = 1}^{p - 1} {{2^{2a}} - \frac{1}{2}(\left( {\frac{2}{p}} \right){2^p} + 1){G_p}}  + 1(\bmod~{2^N} - 1)\\
	& \equiv (\frac{{{2^p} + 1}}{2})(\frac{{{2^N} - 1}}{3} - 1) - \frac{1}{2}(\left( {\frac{2}{p}} \right){2^p} + 1){G_p} + 1(\bmod~{2^N} - 1)\\
	& \equiv ( - \frac{{{2^p} + 1}}{2} + 1) + (\frac{{{2^p} + 1}}{2})(\frac{{{2^N} - 1}}{3}) - \frac{1}{2}(\left( {\frac{2}{p}} \right){2^p} + 1){G_p}(\bmod~{2^N} - 1).
	\end{align*}

(2) By the definition of $G_p$, we know that
	\begin{align*}
	    G_p^2 & = \sum\limits_{x,y=1}^{p - 1} {\left( {\frac{{xy}}{p}} \right)} {4^{x + y}}\\
	& \equiv \sum\limits_{x,t=1}^{p - 1} {\left( {\frac{t}{p}} \right)} {4^{x(1 + t)}}(\bmod~{2^N} - 1)\\
	& \equiv \left( {\frac{{ - 1}}{p}} \right)(p - 1) + \sum\limits_{t = 1}^{p - 2} {\left( {\frac{t}{p}} \right)} \sum\limits_{x = 1}^{p - 1} {{4^{x(1 + t)}}} (\bmod~{2^N} - 1)\\
	& \equiv \left( {\frac{{ - 1}}{p}} \right)(p - 1) + \sum\limits_{t = 1}^{p - 2} {\left( {\frac{t}{p}} \right)} \sum\limits_{x = 1}^{p - 1} {{4^x}} (\bmod~{2^N} - 1)\\
    & \equiv \left( {\frac{{ - 1}}{p}} \right)(p - 1) + (\sum\limits_{t = 1}^{p - 1} {\left( {\frac{t}{p}} \right)}  - \left( {\frac{{p - 1}}{p}} \right))\sum\limits_{x = 1}^{p - 1} {{4^x}} (\bmod~{2^N} - 1)\\
        	\end{align*}
	\begin{align*}
	& \equiv \left( {\frac{{ - 1}}{p}} \right)(p - 1) - \left( {\frac{{ - 1}}{p}} \right)(\frac{{{4^p} - 1}}{3} - 1)(\bmod~{2^N} - 1)\\
	& \equiv \left( {\frac{{ - 1}}{p}} \right)(p - \frac{{{2^N} - 1}}{3})(\bmod~{2^N} - 1).
	\end{align*}
\qed

From formula(7), we can abtain ${\varphi _2}(s) = {\log _2}(\frac{{{2^N} - 1}}{{\gcd ({2^N} - 1,S(2))}})$. Since $N = 2p$, we have ${2^N} - 1 = ({2^p} + 1)({2^p} - 1)$ and $\gcd ({2^p} + 1,{2^p} - 1) = \gcd ({2^p} + 1,2) = 1$. Therefore, $\gcd (S(2),{2^N} - 1)  = \gcd (S(2),{2^p} + 1) \cdot \gcd (S(2),{2^p} - 1)$, where $\gcd (S(2),{2^p} + 1)$ and $\gcd (S(2),{2^p} - 1)$ are odd.

Then, $\gcd (S(2),{2^p} + 1)$ and $\gcd (S(2),{2^p} - 1)$ will be determined in the follows.

\begin{lemma}
 Let $p$ be an odd prime, then,
 \begin{equation*}
\gcd (S(2),{2^p} + 1) = 1.
\end{equation*}
\end{lemma}

\textbf{\emph{Proof}}
Let $l$ be a prime divisor of $\gcd (S(2),{2^p} + 1)$. Since $S(2) \equiv 0(\bmod~{l})$ and ${2^p} + 1 \equiv 0(\bmod~{l})$,  $S(2) \equiv {2^p} + 1 \equiv 0(\bmod~{l})$. From lemma 9(1) and $l|{2^p} + 1$, we have
\begin{equation}
S(2) \equiv 1 - \frac{1}{2}( - \left( {\frac{2}{p}} \right) + 1){G_p}(\bmod~{l}).
\end{equation}
Since ${2^p} + 1 \equiv 1(\bmod~2)$, we have $l \ge 3$.
First, we consider the case that $l = 3$. In this case, ${2^p} + 1 \equiv {( - 1)^p} + 1 \equiv 0(\bmod~3)$, and
\begin{equation*}
G_p \equiv \sum\limits_{a = 1}^{p - 1} {\left( {\frac{a}{p}} \right)} {2^{2a}} \equiv \sum\limits_{a = 1}^{p - 1} {\left( {\frac{a}{p}} \right)}  \equiv 0(\bmod~3).
\end{equation*}

Then from formula(13), we know that $S(2) \equiv 1(\bmod~{l})$, which contradicts to $S(2) \equiv 0(\bmod~{l})$. Therefore, $3$ is not a prime divisor of $\gcd (S(2),{2^p} + 1)$.

Then we assume $l \ge 5$. From formula(13), we know that
\begin{equation}
0 \equiv S(2) \equiv 1 - \frac{1}{2}( - \left( {\frac{2}{p}} \right) + 1){G_p}(\bmod~l).
\end{equation}
and by lemma 9(2), we have $G_p^2 \equiv \left( {\frac{{ - 1}}{p}} \right)p(\bmod~{l})$. If $p \equiv  \pm 1(\bmod~8)$, then $\left( {\frac{2}{p}} \right) = 1$, according to formula(14), $0 \equiv 1(\bmod~{l})$,  it is a contradiction. If $p \equiv  \pm 3(\bmod~8)$ , then $\left( {\frac{2}{p}} \right) =  - 1$, by the formula(14), ${G_p} \equiv 1(\bmod~{l})$, therefore, $1 \equiv G_p^2 \equiv \left( {\frac{{ - 1}}{p}} \right)p(\bmod~{l})$, then $l|p + 1$ or $l|p - 1$. On the other hand, ${2^p} \equiv  - 1(\bmod~{l})$, that is, ${2^{2p}} \equiv 1(\bmod~{l})$, it means the order of $2$ module $l$ is $2p$. According to the Fermat's little theorem, ${2^{l - 1}} \equiv 1(\bmod~{l})$, therefore, $2p|l - 1$. Moreover, $2p \le l - 1 \le  p$, which is also a contradiction.

In summary, $\gcd (S(2),{2^p} + 1) = 1$.
\qed

\begin{lemma}
Let $N = 2p$, and $p$ be an odd prime, then,
\begin{equation*}
\gcd (S(2),{2^p} - 1) = 1.
\end{equation*}

\end{lemma}

\textbf{\emph{Proof}}
Let $l$ be a prime divisor of $\gcd (S(2),{2^p} - 1)$, then ${2^p} - 1 \equiv 0(\bmod~{l})$. According to
\begin{align*}
	    S(2) & = \sum\limits_{i \in D_1^{(2p)}} {{2^i}}  + \sum\limits_{i \in 2D_1^{(p)}} {{2^i}}  + 1 \\
	& \equiv \sum\limits_{i \in D_1^{(2p)}} 1  + \sum\limits_{i \in 2D_1^{(p)}} 1  + 1(\bmod~2)\\
	& \equiv 1(\bmod~2),
	\end{align*}
therefore, $l \ge 3$. Since ${2^p} \equiv 1(\bmod~{l})$ and $l \ge 3$, we know that the order of $2$ module $l$ is $p$. According to the Fermat's little theorem, ${2^{l - 1}} \equiv 1(\bmod~{l})$, therefore $p|l - 1$. On the other hand, from lemma 9, we know that
$$S(2) \equiv 1 - \frac{1}{2}( - \left( {\frac{2}{p}} \right) + 1){G_p}(\bmod~{l}),$$
and
$$G_p^2 \equiv \left( {\frac{{ - 1}}{p}} \right)p(\bmod~{l}).$$
If $p \equiv  \pm 1(\bmod~8),$  then $\left( {\frac{2}{p}} \right) =  1$, moreover, $0 \equiv 1(\bmod~{l})$, it is a contradiction. If $p \equiv  \pm 3(\bmod~8)$, then $\left( {\frac{2}{p}} \right) = - 1$, then $1 \equiv {G_p}(\bmod~{l})$ and $G_p^2 \equiv \left( {\frac{{ 1}}{p}} \right)p(\bmod~{l})$. Therefore, $l|\frac{{p + 1}}{2}$ or $l|\frac{{p - 1}}{2}$, then $p \le l - 1 \le \frac{1}{2}(p + 1) - 1$, which is also a contradiction.

In summary, $\gcd (S(2),{2^p} - 1) = 1.$
\qed

\begin{theorem}
Let $\{ {s_i}\} _{i = 0}^{N - 1}$ be the binary sequences with period $2p$ defined by formula(1). Then the 2-adic complexity of $\{ {s_i}\} _{i = 0}^{N - 1}$ is
$${\varphi _2}(s) = {\log _2}({2^N} - 1).$$
\end{theorem}

\textbf{\emph{Proof}}
From lemma 10 and lemma 11,
$$\gcd (S(2),{2^N} - 1) = \gcd (S(2),{2^p} + 1) \cdot \gcd (S(2),{2^p} - 1) = 1.$$
Therefore,
$${\varphi _2}(s) = {\log _2}(\frac{{{2^N} - 1}}{{\gcd (S(2),{2^N} - 1)}}) = {\log _2}({2^N} - 1).$$
\qed

Finally, we give examples to illustrate Theorem 2.

\textbf{\emph{Example} 7}
Let $ p = 5,g = 3. $ Then,
$$ D_0^{(2p)} = \{ 1,9\} , D_1^{(2p)} = \{ 3,7\} , 2D_0^{(p)} = \{ 2,8\} ,  2D_1^{(p)} = \{ 4,6\} .$$
The corresponding generalized cyclotomic binary sequence of period $10$ is as follows:
$$s = 1001101100.$$

Then, $S(2) = \sum\limits_{i = 0}^9 {{s_i}{2^i}}  = 1 \cdot {2^0} + 0 \cdot {2^1} + 0 \cdot {2^2} + 1 \cdot {2^3} + 1 \cdot {2^4} + 0 \cdot {2^5} + 1 \cdot {2^6} + 1 \cdot {2^7} + 0 \cdot {2^8} + 0 \cdot {2^9} = 217 = 7 \cdot 31.$

For ${2^N} - 1 = {2^{10}} - 1 = 1023 = 3 \cdot 11 \cdot 11,$ ${\varphi _2}(s) = {\log _2}(\frac{{{2^N} - 1}}{{\gcd (S(2), {2^N} - 1)}}) = {\log _2}(\frac{{3 \cdot 11 \cdot 11}}{{\gcd (7 \cdot 31, 3 \cdot 11 \cdot 11)}}) = {\log _2}({2^{10}} - 1).$ The results are consistent with theorem 2.

\textbf{\emph{Example} 8}
Let $ p = 7,g = 3.$ Then,
$$ D_0^{(2p)} = \{ 1,9,11\} , D_1^{(2p)} = \{ 3,5,13\} , 2D_0^{(p)} = \{ 2,4,8\} ,  2D_1^{(p)} = \{6,10,12\} .$$
The corresponding generalized cyclotomic binary sequence of period $14$ is as follows:
$$s = 10010110001011.$$

Then, $S(2) = \sum\limits_{i = 0}^{13} {{s_i}{2^i}}  = 1 \cdot {2^0} + 0 \cdot {2^1} + 0 \cdot {2^2} + 1 \cdot {2^3} + 0 \cdot {2^4} + 1 \cdot {2^5} + 1 \cdot {2^6} + 0 \cdot {2^7} + 0 \cdot {2^8} + 0 \cdot {2^9} + 1 \cdot {2^{10}} + 0 \cdot {2^{11}} + 1 \cdot {2^{12}} + 1 \cdot {2^{13}} = 13417.$

For ${2^N} - 1 = {2^{14}} - 1 = 16383 = 3 \cdot 43 \cdot 127,$ ${\varphi _2}(s) = {\log _2}(\frac{{{2^N} - 1}}{{\gcd (S(2),{2^N} - 1)}}) = {\log _2}(\frac{{{2^{14}} - 1}}{{\gcd (13417,16383)}}) = {\log _2}(\frac{{{2^{14}} - 1}}{{\gcd (13417,3 \cdot 43 \cdot 127)}}) = {\log _2}({2^{14}} - 1).$ The results are consistent with theorem 2.

\section{Conclusion}
\label{sec:1}
This paper determined the linear complexity, minimal polynomial and 2-adic complexity of a class of generalized cyclotomic binary sequences with period $2p$, which are constructed base on generalized cyclotomy. The results show that the linear complexity and the 2-adic complexity of the considered sequences are reach maximum when $p\equiv \pm 1(\bmod~8)$ over extension field. And the minimal value of the linear complexity of these sequences is equal to $p + 1$, which is greater than $p$, the half of the period of these sequences. In addition, by the literature \cite{23} method, such sequences have low-value autocorrelation. That is, such sequences has low-value autocorrelation, the highest linear complexity and the highest 2-adic complexity when $p\equiv \pm 1(\bmod~8)$. Therefore, the class of sequences can be viewed as enough good for pseudorandom sequences in terms of cryptographyic criterions.

\section{Acknowledgement}
The work of Y. Wang was supported by the National Natural Science Foundation of China under Grant 61902304. The work of Z. Heng was supported by the National Science Foundation of China under Grant 11901049.

\end{document}